\documentclass[letterpaper, 10 pt,one column, conference]{ieeeconf}  % Comment this line out if you need a4paper
\usepackage{graphicx,graphics}
\usepackage{amsmath,amsfonts,mathrsfs,amssymb,dsfont}
\usepackage{booktabs}
\usepackage{cite}
\usepackage{sidecap}
\sidecaptionvpos{figure}{l}
\usepackage{bbm}
\usepackage{color}
\usepackage{arydshln}

\IEEEoverridecommandlockouts                              % This command is only needed if 
\title{\LARGE \bf
Control design and analysis of a stochastic network control system 
}

\author{Mohammad Soltani$^{1}$, Abhyudai Singh$^{2}$% <-this % stops a space
\thanks{$^{1}$M. Soltani is with Department of Electrical and Computer Engineering, University of Delaware, Newark, DE USA 19716.
{\tt\small msoltani@udel.edu}}%
\thanks{$^{2}$A. Singh is with the Department of Electrical and Computer Engineering, Biomedical Engineering, Mathematical Sciences, Center for Bioinformatics and Computational Biology, University of Delaware, Newark, DE USA 19716.
{\tt\small absingh@udel.edu}}}

\begin{document}

\maketitle
\thispagestyle{empty}
\pagestyle{empty}

\begin{abstract}
%A Network Control System (NCS) consists of control components that interact over a shared network.
%In this paper, we study a NCS in which data transmission between controller and plant occurs at random times. 
%We consider a minimal NCS that consists of a plant, and a controller. We further assume that noise in the system arises from two sources: i) external disturbance that affects the plant dynamics, ii) noise acquired in the transmission of the data over the communication network. 
%For this NCS, we investigate how the statistical properties of the data transmission times affects the state of the system, and under what conditions the system is stable. A further question of interest is designing a control law could steer the states of the system to reach to a desired mean and variance. To this end, we derive exact dynamics of the first two moments of the system , and use them to derive the stability conditions. Finally, we demonstrate our results using different examples, and show that randomness in the data transmission times can reduce the variability contributed from disturbance.

A Network Control System (NCS) consists of control components that interact with the plant over a shared network. The system dynamics of a NCS could be subject to noise arising from randomness in the times at which the data is transmitted over the network, corruption of the transmitted data by the communication network, and external disturbances that might affect the plant. A question of interest is to understand how the statistics of the data transmission times affects the system dynamics, and under what conditions the system is stable. Another related issue is designing a controller that meets desired performance specifications (e.g., a specific mean and variance of the system state). Here, we consider a minimal NCS that consists of a plant and a controller, and it is subject to random transmission times, channel corruption and external disturbances.  We derive exact dynamics of the first two moments of the system, and use them to derive the stability conditions of the system. We further design a control law that steers the system to a desired mean and variance. Finally, we demonstrate our results using different examples, and show that under some specific conditions, randomness in the data transmission times can even reduce the variability contributed from disturbance.

\end{abstract}
%%%%%%%%%%%%%%%%%%%%%%%%%%%%%%%%%%%%%%%%%%%%%%%%%%%%%%%%%%%%%%%%%%%%%%%%%%%%%%%%
\section{INTRODUCTION}
Advent of modern information and communication technologies has provided the ability to design control systems whose elements are located at remote distances. In such Network Control Systems (NCSs), different components of a feedback loop communicate through a shared network \cite{hes_05,HespanhaMar04,BohacekHespanhaLeeObraczkaJun03}. The dynamics of a NCS is affected by noise arising from different sources. For example, different components of a NCS are typically designed to be facultative so as to reduce the energy consumption and use the shared network resources efficiently \cite{lin07,gad14}. Since the availability of the network at a given time depends on whether or not other components of the NCS are using it, the times at which a specific device is able to communicate is inherently random. The transmitted data can further be corrupted by the noise in the communication channel. Moreover, the plant dynamics could be subject to external disturbances. Collectively, these noise sources hamper the overall performance of the system.  Previous works on NCS have dealt with designing control strategies to improve system performance. In particular, for NCS with limited resources, different strategies are desired that reduce the number of times a control component needs to communicate over the network. Examples of such strategies include event-triggered control, where transmission happens when the state of the system meets a certain condition \cite{tab07,hjt12,ank16}; and self-triggered control, where the state of the system at the time of a transmission is used to determine the next transmission time \cite{ant10,wal09,aah17}. Further, motivated by specific requirements of these systems, new network protocols are designed \cite{dan08}.

Here, we analyze the effect of various noise sources on stability of the system, and also design a control law that can drive the system to a desired state. We consider a NCS whose states are modeled as Stochastic Differential Equations (SDEs). These SDEs naturally capture the disturbances in the system. For example, if the disturbance is state independent (dependent) then the corresponding drift and diffusion terms of the SDEs could be taken as state independent (dependent). To capture the randomness in the transmission times, we model them using a renewal process. We provide the necessary and sufficient conditions on stability of first- and second-order moments of the states of NCS under consideration. Further, we quantify the contribution of each noise source to the states of the system. We show that the exact results combined with stability conditions can be used to design controllers to meet desired performance criterion. Finally ,we demonstrate our method via different examples. An interesting observation is that while rare and randomly transmission times are expected to increase noise in a system, these might even reduce the noise in the system in some parameter regimes. 

The rest of this paper is organized as following: In Section II network control systems are formulated. Section III provides exact moments of these systems, and in Section IV and V examples of such systems are investigated. Finally conclusion and direction of future works are presented in Section VI.

\begin{figure*}[!th]
	\centering
	{\includegraphics[width=0.85\textwidth]{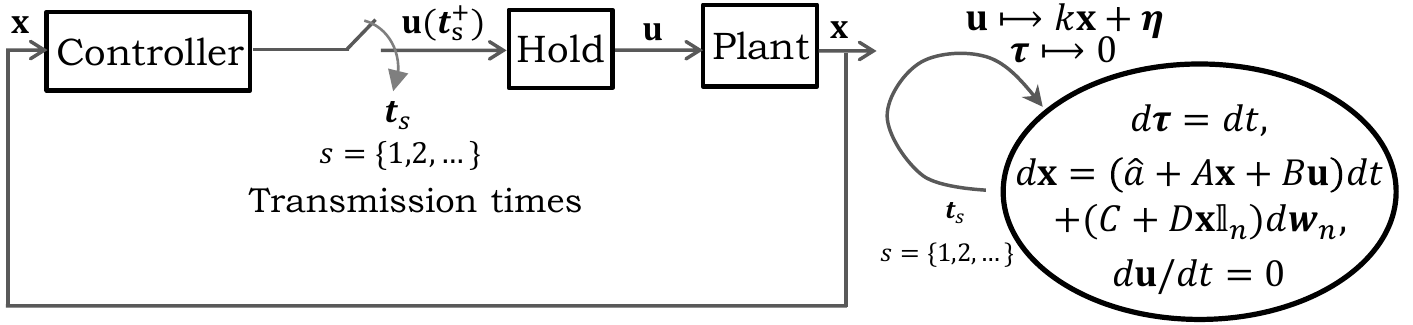}}
	\caption{{\bf Model schematic of a stochastic NCS model}. \textit{Left}: The controller is far from the plant hence the feedback loop is connected through a network. In between transmission times, plant uses the previous control law which is maintained in hold. Any time that connection occurs, the hold reads the new control law which is calculated based on the current values of the states of the system. \textit{Right}: Mathematical representation of network control system. Resets here are the times in which connection occurs. Any time that a transmission occurs, new control law is applied to the plant. However due to measurements error, an extra term $\eta$ is added to the system.}
\end{figure*}

\section{Stochastic control of linear systems}
Let the states of the system $\bold x \in \mathbb{R}^{n \times 1}$ evolve according to the 
following stochastic differential equation
\begin{equation}
d\bold x= \left(\hat{a}+A \bold x(t)+B \bold u(t)\right)dt + (C+D\bold{x}(t)\mathbbm{1}_n)d \boldsymbol w_n ,
		\label{dynamicsSDE}	
		\end{equation}
where $\bold u (t)\in \mathbb{R}^{m \times 1}$ denotes the controller; $\hat{a}\in \mathbb{R}^{n\times 1}$ is a constant vector; $A\in \mathbb{R}^{n\times n}$, $B \in \mathbb{R}^{ n \times m}$, $C\in \mathbb{R}^{n\times n}$ and $D \in \mathbb{R}^{ n \times n}$ are constant matrices;  and $\mathbbm{1}_n$ is a $1 \times n$ unit matrix. In addition, $\boldsymbol w_n$ is an $n$-dimensional Weiner process that satisfies the following
\begin{equation}
\langle d \boldsymbol w_n \rangle = 0, \ \   \langle d \boldsymbol w_n  d \boldsymbol w_n ^\top \rangle = I_n dt.
\end{equation}
Here $I_n$ is an $n \times  n$ Identity matrix, and the symbol $\langle \ \rangle$ denotes the expected value. While the first part in the right-hand side of \eqref{dynamicsSDE} determines dynamics of the plant, the second part represent the contribution of disturbance to the system. The state-independent disturbance is modeled through $C$, and $D\bold{x}(t)\mathbbm{1}_n$ represents state-dependent disturbance. 	 

The controller connects to the plant at random times $\boldsymbol{t}_s, \ s\in \{1,2,\ldots\}$ as shown in Fig. 1. We assume that the time intervals between transmission times
\begin{align}
\boldsymbol T_s  \equiv   \boldsymbol t_s - \boldsymbol  t_{s-1}\label{pdf of T}
\end{align}
are independent and identically distributed (iid) random variables which follow a generic, continuous probability density function $f$ with support over non-negative real line. Whenever the controller communicates with the plant, the control law reset as
\begin{equation}
\bold u (\boldsymbol t_s^+) \mapsto K \bold x(\boldsymbol  t_s^-)+\boldsymbol \eta, \ K\in  \mathbb{R}^{m\times n} ,\label{reset}
\end{equation}
where $\boldsymbol t_s^-$ and $\boldsymbol t_s^+$ denote the time just before and after a transmission, respectively. The matrix $K$ here consists of controller gains and the term $\boldsymbol \eta$ denotes the noise of the communication channel. We assume that $\boldsymbol \eta$ is a vector of zero-mean noise terms and $\langle \boldsymbol \eta \boldsymbol \eta^\top \rangle = \Sigma \in \mathbb{R}^{m\times m} $, where $\Sigma$ is a diagonal matrix. The control law is assumed to remain constant in between the consecutive transmission events, i.e.,
\begin{equation}
\frac{d\bold u}{dt} =0. \label{control law}
\end{equation}

In order to obtain a mathematically tractable model, we introduce a timer $\boldsymbol \tau$ that measures the time since the last transmission. The timer increases with time in between transmissions, and resets to zero whenever the transmissions occur. Let the 
probability that a transmission occurs in an infinitesimal time interval, $(t,t+dt]$, be $h(\boldsymbol \tau)dt$. Then, we have that
\begin{equation} \label{hr}
h( \tau ) \equiv   \frac{f(\tau)}{1-\int_{y=0}^{\tau}f(y)dy}.
\end{equation}
Alternatively, the duration between events $\boldsymbol T_s$ follows a probability density function $f$ given by
\begin{equation}
\boldsymbol T_s \sim  f(\tau) = h(\tau )  {\rm e}^{-\int_0^{\tau} h(y) d y}
\end{equation}
\cite{Ross20109,ehp00}, and the timer follows the following steady-state probability density function 
\begin{equation}
\boldsymbol \tau \sim  p(\tau) = \frac{1}{\langle \boldsymbol T_s\rangle} {\rm e}^{-\int_0^{\tau} h(y) d y}, \label{prob. tau}
\end{equation}
where $\langle \boldsymbol T_s\rangle$ is the mean time interval in between transmissions \cite{vss16}. Modeling the timing of transmissions through the timer enables us to investigate statistical moments of such systems as described in the next section. 

\section{Statistical moments of NCS}
We start our analysis by introducing a new vector that  contains both states and controller $\bold y= [\bold x \ \bold u]^\top\in \mathbb{R}^{n+m\times 1}$. Dynamics of this vector is obtained from \eqref{dynamicsSDE} and \eqref{control law} as
\begin{equation}
d\bold y= \left(\hat{a}_y+A_y \bold y (t)\right)dt + (C_y+D_y\bold{y}(t)\mathbbm{1}_{n+m})d \boldsymbol w_{n+m} ,
		\label{dynamicsSDE2}	
\end{equation}
where 
	\begin{equation}
 \hat{a}\equiv  \left[\begin{array}{c}
	\hat{a}\\ \hdashline[2pt/2pt]0 
	\end{array}\right],  \ \  A_y\equiv  \left[\begin{array}{c;{2pt/2pt}c}
	A & B\\ \hdashline[2pt/2pt]0 & 0 
	\end{array}\right],\ \ 
C_y\equiv  \left[\begin{array}{c;{2pt/2pt}c}
	C & 0\\ \hdashline[2pt/2pt]0 & 0 	\end{array}\right], D_y\equiv  \left[\begin{array}{c;{2pt/2pt}c}
	D & 0\\ \hdashline[2pt/2pt]0 & 0 	\end{array}\right], d \boldsymbol w_{n+m} \equiv  \left[\begin{array}{c}
	d \boldsymbol w_{n} \\ \hdashline[2pt/2pt]0  	\end{array}\right].
	\end{equation}
Furthermore, at any time when the controller transmits a new control law, the states of $\bold y$ change as
\begin{align}
&\langle \bold{y} (\boldsymbol{t}_s^+)\rangle=  J\bold{y}(\boldsymbol{t}_s^-), \ J \equiv \left[\begin{array}{c;{2pt/2pt}c}
	I_n & 0\\ \hdashline[2pt/2pt] K & 0 	\end{array}\right],\label{conditional x}
\end{align}
where we used the fact that the states of the system will not change during the events thus $\langle \bold x( \boldsymbol t_s^+) \rangle=\bold x( \boldsymbol t_s^-)$.

\subsection{Mean of NCS}
In Appendix A, we show that the steady-state mean of vector $\bold y$ is bounded, if and only if the expected value 
	\begin{equation}
 \left\langle  {\rm e}^{ A_y \boldsymbol T_s} \right \rangle =  \int_0^{\infty}f(\tau) e^{A_y\tau }d \tau 
	\end{equation}
exists and all of the eigenvalues of the matrix $J_y \left\langle  {\rm e}^{ A_y \boldsymbol T_s} \right \rangle$ are inside the unit cycle, i.e.
\begin{equation}
\big \vert \text{eig}\left(J_y \langle e^{A_y \boldsymbol T_s} \rangle  \right) \big \vert <1 . \label{eigs0}
\end{equation}
In this limit, the steady-state mean of vector $\overline{\langle \bold x \rangle} =\lim_{t\to\infty}\langle \bold x \rangle$ is 
\begin{equation}
\begin{aligned}
\overline{\langle {\bold{y}} \rangle} =  & \left\langle {\rm e}^{ A_y\boldsymbol \tau} \right \rangle \left(I_{n +m}-J _y\left\langle {\rm e}^{ A_y\boldsymbol T_s} \right \rangle   \right)^{-1}J_y\left \langle  {\rm e}^{ A_y\boldsymbol T_s} \int_0^{\boldsymbol T_s}  {\rm e}^{ -A_yr} \hat{a}_y dr  \right \rangle  + \left \langle {\rm e}^{ -A_y \boldsymbol \tau}  \int_0^{\boldsymbol \tau}  {\rm e}^{ -A_yr} \hat{a}_y dr  \right \rangle , 
\label{mean of x general}	
\end{aligned}
\end{equation} 
where the vector
\begin{equation}
\left \langle  {\rm e}^{ A_y\boldsymbol T_s} \int_0^{\boldsymbol T_s}  {\rm e}^{ -A_yr} \hat{a} dr  \right \rangle  =  \int_0^{\infty} f(\tau )\left( {\rm e}^{ A_y\tau } \int_0^{\tau }  {\rm e}^{ -A_yr} \hat{a}_y dr \right)  d \tau 
\end{equation}
is obtained by taking expected value with respect to $\boldsymbol T_s$, and 
	\begin{align}
	& \left\langle  {\rm e}^{ A\boldsymbol \tau } \right \rangle =  \int_0^{\infty}p(\tau) e^{A\tau }d \tau, \ \  \left \langle  {\rm e}^{ A\boldsymbol \tau } \int_0^{\boldsymbol \tau }  {\rm e}^{ -Ar} \hat{a} dr  \right \rangle  =  \int_0^{\infty} p(\tau )\left( {\rm e}^{ A\tau } \int_0^{\tau }  {\rm e}^{ -Ar} \hat{a} dr \right)  d \tau 
	\end{align}
are derived by taking expected value with respect to $\boldsymbol \tau$.
In the next part, we provide an approach to derive the second-order moments of the system.

\subsection{Second-order moments of NCS}
Our strategy is to transform the second-order moments to a similar form as in \eqref{dynamicsSDE2} and \eqref{conditional x}. To that end, we introduce a new vector 
 \begin{equation}\boldsymbol \mu=[\bold x \ \bold u \ {\rm vec}(\bold x \bold x^\top)  \ {\rm vec}(\bold x \bold u^\top)  \ {\rm vec}(\bold u \bold u^\top)  ],\end{equation}
where ${\rm vec}(M)$ denotes the vector transformation of a matrix $M$. Vectorization is transforming a matrix into a column vector by putting all the columns of the matrix into a vector subsequently. In appendix B, we show that dynamics of $\boldsymbol \mu$ in between the events is given by 
 \begin{equation}
\dot{\boldsymbol \mu}= \hat{a}_\mu + A_\mu \boldsymbol \mu , \label{mu dynamic0}
\end{equation}
where
\begin{equation}\label{au}
\begin{aligned}
\hat{a}_\mu\equiv &  \left[\begin{array}{c}
\hat{a}\\ \hdashline[2pt/2pt]0\\ \hdashline[2pt/2pt] \small{ {\rm vec}  \left( CC^\top\right)} \\ \hdashline[2pt/2pt]0\\ \hdashline[2pt/2pt]0
\end{array}\right],  
A_\mu \equiv   \left[\begin{array}{c;{2pt/2pt}c;{2pt/2pt}c;{2pt/2pt}c;{2pt/2pt}c}
	A & B & 0 & 0& 0\\ \hdashline[2pt/2pt]0 & 0 & 0 & 0&0 \\ \hdashline[2pt/2pt] M_1& 0&  M_2&M_3 & 0  \\ \hdashline[2pt/2pt] 0& I_m \otimes \hat{a}&  0&I_m \otimes A & I_m \otimes B \\ \hdashline[2pt/2pt] 0&0&  0&0& 0	\end{array}\right],\\ 
M_1\equiv &I_n \otimes \hat{a} + \hat{a}  \otimes I_n+D \otimes C  \mathbbm{1}_{n}+ C^\top  \mathbbm{1}_{n} \otimes D, \\
M_2 \equiv &I_n \otimes A+  A \otimes I_n +n D\otimes D,     \\ 
M_3 \equiv & I_n \otimes B+  B \otimes I_n,
\end{aligned}
\end{equation}
and $\otimes$ denotes the Kronecker product. Furthermore, at the time of transmission, the states of the vector $\boldsymbol\mu$ change as (see Appendix C)
\begin{equation}
\langle \boldsymbol \mu(\boldsymbol{t}_s^+)  \rangle =   J_\mu \boldsymbol \mu(\boldsymbol{t}_s^-) + R_\mu,  \ \ 
J_\mu \equiv    \left[\begin{array}{c;{2pt/2pt}c;{2pt/2pt}c;{2pt/2pt}c;{2pt/2pt}c}
	I & 0 & 0 & 0& 0\\ \hdashline[2pt/2pt]0 &  K & 0 & 0&0 \\ \hdashline[2pt/2pt] 0& 0& I&0& 0  \\ \hdashline[2pt/2pt] 0& 0&  K \otimes I_n &0 & 0\\ \hdashline[2pt/2pt] 0&0&  K \otimes K  &0& 0	\end{array}\right], \ \
R_\mu \equiv  \left[\begin{array}{c}
0\\ \hdashline[2pt/2pt]0\\ \hdashline[2pt/2pt] 0\\ \hdashline[2pt/2pt]0\\ \hdashline[2pt/2pt]\Sigma
\end{array}\right]. \label{reset 2}
\end{equation}

The deterministic dynamics in \eqref{mu dynamic0} and stochastic resets in \eqref{reset 2} are similar to those in \eqref{dynamicsSDE2} and \eqref{conditional x}. Therefore, with a similar analysis as in Appendix A, the steady-state mean of vector $\boldsymbol \mu$ is bounded, if and only if all of the eigenvalues of the matrix $J_\mu \left\langle  {\rm e}^{ A_\mu \boldsymbol T_s} \right \rangle$ are inside the unit cycle. In this limit $\overline{\langle \boldsymbol \mu \rangle} =\lim_{t\to\infty}\langle \boldsymbol \mu \rangle$ is given by
	\begin{equation}
	\begin{aligned}
	\overline{\langle {\boldsymbol{\mu}} \rangle} =  & \left\langle {\rm e}^{ A_\mu\boldsymbol \tau} \right \rangle \left(I_{n+m+n^2+m^2+nm} -J_\mu\left\langle {\rm e}^{ A_\mu \boldsymbol T_s} \right \rangle   \right)^{-1} \left(J_\mu \left \langle  {\rm e}^{ A_\mu\boldsymbol T_s} \int_0^{\boldsymbol T_s}  {\rm e}^{ -A_\mu r} \hat{a}_\mu dr  \right \rangle+R_\mu\right)  + \left \langle {\rm e}^{ -A_\mu \boldsymbol \tau}  \int_0^{\boldsymbol \tau}  {\rm e}^{ -A_\mu r} \hat{a}_\mu dr  \right \rangle .
	\label{mean of x2}	
	\end{aligned}
	\end{equation} 
The last term in the above expression can be computed by taking expected value with respect to $\boldsymbol T_s$
\begin{equation}\begin{aligned}
&	\left \langle  {\rm e}^{ A_\mu\boldsymbol T_s} \int_0^{\boldsymbol T_s}  {\rm e}^{ -A_\mu r} \hat{a} dr  \right \rangle  =  \int_0^{\infty} f(\tau )\left( {\rm e}^{ A_\mu\tau } \int_0^{\tau }  {\rm e}^{ -A_\mu r} \hat{a}_\mu dr \right)  d \tau .
	\end{aligned}
Likewise, we get	
	\end{equation}
		\begin{align}
 \left\langle  {\rm e}^{ A_\mu \boldsymbol \tau } \right \rangle =  \int_0^{\infty}p(\tau) e^{A_\mu \tau }d \tau,  \ \  \left \langle  {\rm e}^{ A_\mu \boldsymbol \tau } \int_0^{\boldsymbol \tau }  {\rm e}^{ -A_\mu r} \hat{a}_\mu dr  \right \rangle  =  \int_0^{\infty} p(\tau )\left( {\rm e}^{ A_\mu\tau } \int_0^{\tau }  {\rm e}^{ -A_\mu r} \hat{a}_\mu dr \right)  d \tau 
		\end{align}
by taking expected value with respect to $\boldsymbol \tau$. Mean of $\boldsymbol \mu$ contains all the second-order moments of the vector $\bold x$.

\subsection{Control design from steady-state moments}
We can design the controller for having desired steady-state mean in \eqref{mean of x general} by choosing $K$.
This equation combined with stability condition tells whether the desired mean is possible or not, i.e., as long as the eigenvalues of the matrix $J_\mu \left\langle  {\rm e}^{ A_\mu \boldsymbol T_s} \right \rangle$ are inside the unit cycle, we can design the matrix of control gains $K$ to have the desired steady-state mean of $\bold x$ and finite second-order moments.

Moreover, by having exact solutions of the second-order moments, we can also design a controller that minimizes the steady-state variance of the system. We can set up this problem as an optimization problem 
\begin{subequations}
	\begin{align}
	&\text{Minimize}_K \left({\rm vec}(\overline{\langle \bold x \bold x^\top \rangle}) - {\rm vec}(\overline{\langle \bold x \bold \rangle} \ \overline{\langle \bold x \bold \rangle}^\top ) \right),\\
	&\text{Subject to }  \overline{\langle \bold x \bold \rangle}= {\rm Constant} .
	\end{align}
\end{subequations}

It may not always be feasible to obtain a control law that meets all desired specifications. This happens when the degrees of freedom in the controller is less than number of mean and variance terms. We illustrate this by an example in the next section wherein one can design the controller to achieve to a specific mean, but then controller is unable to change the variance to a desired level.

\begin{figure}[!h]
	\centering
	{\includegraphics[width=0.97\textwidth]{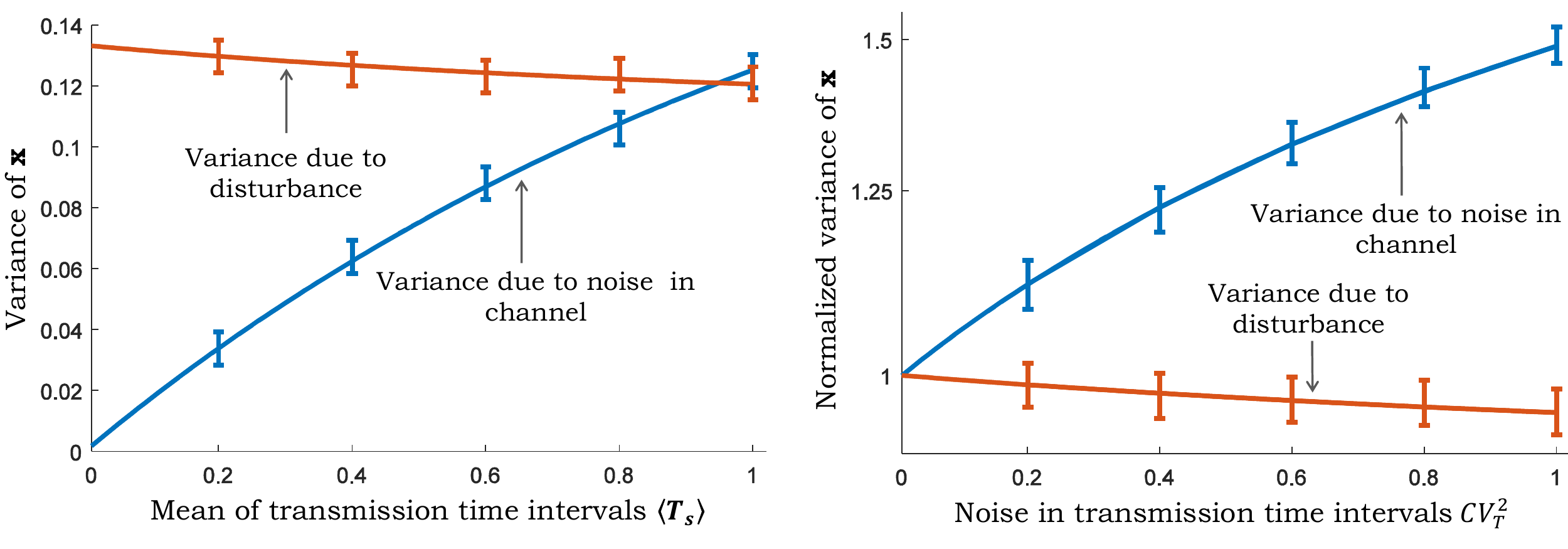}}
	\caption{{\bf Making transmission times more random  can reduce variance in $\bold x$}.
\textit{Left}) Surprisingly by increasing the mean time interval in between the transmissions, noise in $\bold x$ contributed from disturbance can reduce. On the other hand increasing $\boldsymbol T_s$ will increase the noise contributed from noisy channel drastically. It is because the added noise by channel remains in the system for a longer time before getting corrected by a new control. \textit{Right})
Interestingly noisy transmission times also can reduce the variance contributed from disturbance. Hence when noise in channel is small, randomly distributed control sequence can be used to reduce variance in $\bold x$. For this plot we used gamma distributed time intervals and variance of $\bold x$ is normalized to its value at the beginning of the plot. Noise in the inter transmission times is quantified by coefficient of variation squared $CV^2_T =\frac{\langle \boldsymbol T_s^2 \rangle - \langle \boldsymbol T_s \rangle^2 }{\langle \boldsymbol T_s \rangle^2}$. 
  The  parameters are selected as $\hat{a}=1$,  $a = -1$, $c=0.45$, $b = 0.5$, $ k = 0.5$, and $\sigma=1$. $95\% $ confidence intervals are obtained by running $1000$ numerical simulations. 
}
\end{figure}

\section{A scalar network control system}
Suppose that the state of a system is governed via a one dimensional SDE as
\begin{equation}
d\bold x= \left(\hat{a}+a \bold x(t)+b \bold u(t)\right)dt + c  \ d \boldsymbol w,
		\label{dynamicsSDE00}	
		\end{equation}
and the control law at the time of resets is
\begin{equation}
\bold u (\boldsymbol t_s^+) \mapsto k \bold x(\boldsymbol  t_s^-)+\boldsymbol \eta, \label{reset000}
\end{equation}
where $\boldsymbol \eta$ is a zero mean noise term with variance $\sigma^2$. 
Based on \eqref{dynamicsSDE2}, the vector $\bold y=[\bold x \ \bold u]^\top $ is governed via 
\begin{equation}
d\bold y= \left(\hat{a}_y+A_y \bold y (t)\right)dt + C_yd \boldsymbol w_2 ,
		\label{dynamicsSDE0}	
\end{equation}
where 
	\begin{equation}
 \hat{a}= \left[\begin{array}{c}
	\hat{a}\\ 0  
	\end{array}\right],  \  A_y= \left[\begin{array}{cc}
	a & b\\ 0 & 0 
	\end{array}\right],
C_y= \left[\begin{array}{cc}
	c & 0\\ 0 & 0 	\end{array}\right].
	\end{equation}
Further, at the time of connection the states reset as \eqref{conditional x} with $J= \left[\begin{array}{cc}
	1& 0\\ - k & 0 	\end{array}\right]$.
Hence, based on \eqref{mean of x general}, the mean of the state $\bold x$ in steady-state is 
\begin{equation} \label{mean}
\overline{\langle {\bold{x}} \rangle} = -\frac{\hat{a}}{a+bk}.
\end{equation}	
Interestingly, the mean of $\bold x$ in steady-state is independent of $\boldsymbol T_s$. This implies that as long as the system is stable, having rare transmissions of the data does not change the mean of the states.

\subsection{Second-order moment}
In order to derive the second-order moments, we define
\begin{equation}
\boldsymbol \mu = [\langle \bold x \rangle \ \langle \bold u \rangle \ \langle \bold x^2 \rangle\ \langle \bold x \bold u \rangle \ \langle \bold u^2 \rangle ].
\end{equation}
For this $\boldsymbol \mu$, $\hat{a}$ and $A_\mu$ in \eqref{au} are defined as 
	\begin{equation}
 a_\mu =   \left[\begin{array}{c}
	\hat{a} \\ 0 \\ c^2\\ 0 \\ 0  \end{array}\right], \ \ A_\mu = \left[\begin{array}{ccccc}
	a & b& 0 & 0 & 0\\ 0 & 0& 0 & 0& 0 \\ 2\hat{a} & 0& 2 a& 0& 0 \\ 2b & \hat{a}& 0 & a& b \\ 0 & 0& 0 & 0& 0
	\end{array}\right]. \ \ 
	\end{equation}
Furthermore, at any time when the transmission occurs over the network, the states of the system reset to
\begin{equation}
\langle \boldsymbol \mu_+(\boldsymbol t_s) \rangle  = J_\mu \boldsymbol \mu(\boldsymbol t_s)+ R_\mu, \label{mu reset0}
\end{equation}
where
\begin{equation}
J_\mu = \left[\begin{array}{ccccc}
	1 & 0& 0 & 0 & 0\\ k& 0& 0 & 0& 0 \\ 0 & 0& 1& 0& 0\\ 0 & 0& k & 0& 0 \\ 0 & 0& k^2 & 0& 0
	\end{array}\right], R_\mu = \left[\begin{array}{c}
	0\\ 0 \\ 0 \\ 0 \\ \sigma^2
	\end{array}\right].
\end{equation}

Using these matrices, the non-zero eigenvalues of $J_\mu\left\langle {\rm e}^{A_\mu \boldsymbol T_s }\right\rangle $ are given by
\begin{equation}
\begin{aligned}
&\text{eig}\left(J_\mu \langle e^{A_\mu \boldsymbol T_s} \rangle  \right)=\left \{ \frac{a\left \langle e^{a \boldsymbol T_s} \right \rangle +b \left \langle e^{a \boldsymbol T_s} \right \rangle  k-b k}{a},
\frac{a\left \langle e^{2 a \boldsymbol T_s} \right \rangle   (a+b k)^2+b k (b k-2 a\left \langle e^{a \boldsymbol T_s} \right \rangle   (a+b k))}{a^2}\right \}. \label{eig}
\end{aligned}
\end{equation}
If these eigenvalues are inside the unit circle, then mean and variance of the state $\bold x$ is given by \eqref{mean} and 
\begin{equation}\footnotesize
\begin{aligned}
& \text{var}(\bold x )= \overline{\langle \bold x^2 \rangle }-\overline{\langle \bold x \rangle }^2=\overbrace{\sigma^2 \frac{b^2 (a \langle \boldsymbol T_s \rangle - \left \langle e^{a\boldsymbol T_s }\right \rangle +1) \left(a^2 (\left \langle e^{a\boldsymbol T_s }\right \rangle+1) \langle \boldsymbol T_s \rangle +2 b k (\left \langle e^{a\boldsymbol T_s }\right \rangle (a \langle \boldsymbol T_s \rangle -1)+1)\right)}{a^3 \langle \boldsymbol T_s \rangle ^2 (a+b k) (a (\left \langle e^{a\boldsymbol T_s }\right \rangle +1)+b (\left \langle e^{a\boldsymbol T_s }\right \rangle -1) k)}}^\text{Fluctuations contributed from noisy channel}\\ 
& + \overbrace{c^2 \frac{\left(-a^4 (\left \langle e^{a\boldsymbol T_s }\right \rangle  +1) \langle \boldsymbol T_s \rangle ^2+a^2 b k \langle \boldsymbol T_s \rangle  (-2 a \left \langle e^{a\boldsymbol T_s }\right \rangle  \langle \boldsymbol T_s \rangle +\left \langle e^{2a\boldsymbol T_s }\right \rangle -1)+2 b^2 k^2 (-a \langle \boldsymbol T_s \rangle +\left \langle e^{a\boldsymbol T_s }\right \rangle  -1) (\left \langle e^{a\boldsymbol T_s }\right \rangle  (a \langle \boldsymbol T_s \rangle -1)+1)\right)}{2 a^3 \langle \boldsymbol T\rangle ^2 (a+b k) (a \left \langle e^{a\boldsymbol T_s }\right \rangle  +1)+b (\left \langle e^{a\boldsymbol T_s }\right \rangle  -1) k)}.}^\text{Fluctuations contributed from randomness in dynamics of the system}
\end{aligned} 
\label{noise}
\end{equation}
Interestingly, in some parameter regimes rare control of system (i.e., transmitting the control law after long times) can result in lower variance of the system. To see this, suppose that dynamics of the system are noisy and, i.e., $c$ is large and fluctuations due to disturbance are dominant. In this limit, rare control of the system can reduce the noise as shown in Fig. 2. Note that the variance of $\bold x$ in this limit is less than
\begin{equation}
-\frac{c^2}{a+bk}
\end{equation}
which is the variance for the case in which controller and plant are connected all the time.
This means that rare transmissions on the network not only saves resources such as bandwidth and energy, but also reduces the variance in $\bold x$.
Moreover, in Fig. 2 we also illustrated the effect of noise in transmission times on variance of $\bold x $. As expected, increasing noise in timing of transmissions increases fluctuations that arises from noise in channel. However, again these noisy transmission times can be used to reduce the effect of fluctuations contributed from disturbance.

It is important to point that the aforementioned scenario just occurs in specific parameter regimes. In most of the cases, noise increases by increasing mean time intervals and randomness in timing of transmissions.

\subsection{Control design}	
For this system, the non-zero eigenvalues of the matrix $J_y \left \langle e^{A \boldsymbol T_s }\right \rangle$ are given by \eqref{eig}. 
Therefore, in order to have finite mean of $\bold x$, it is necessary and sufficient that 
	\begin{align}
\big \vert \frac{a\left \langle e^{a \boldsymbol T_s} \right \rangle +b \left \langle e^{a \boldsymbol T_s} \right \rangle  k-b k}{a} \big \vert<1,\ \ 
 \big \vert \frac{a\left \langle e^{2 a \boldsymbol T_s} \right \rangle   (a+b k)^2+b k (b k-2 a\left \langle e^{a \boldsymbol T_s} \right \rangle   (a+b k))}{a^2}  \big \vert<1, \label{eigen value}
	\end{align} 
where $\vert \ \vert$ denotes the absolute value.
In addition, we can design the controller for having desired mean of the state by choosing $k$ as
\begin{equation}
k=-\frac{\hat{a}+ a\overline{\langle \bold x \rangle} }{b\overline{\langle \bold x \rangle}} \label{K}.
\end{equation}
It should be noted that the designed value of $k$ should satisfy \eqref{eigen value}. Since $k$ is a scalar and we just have one degree of freedom in designing controller,  we can only achieve the desired mean level. 

\subsection{The limit of $a=0$}

As an interesting and easy to follow limit, consider that $a=0$. In this limit, the non-zero eigenvalue of $J_y \left\langle  {\rm e}^{ A_y \boldsymbol T_s} \right \rangle$ is $1+bk \langle \boldsymbol T_s \rangle $. Thus, in order to have a finite mean of $\bold x$, the sign of $k$ should be selected opposite to that of $b$, mean of inter transmission times should be finite, and $\vert 1+bk \langle \boldsymbol T_s \rangle \vert <1$. In this limit, mean of $\bold x$  in steady state is 
\begin{equation}
\overline{\langle {\bold{x}} \rangle} = -\frac{\hat{a}}{bk}.
\end{equation}

Furthermore, to have finite second-order moment, in addition to a finite mean, the following eigenvalue also should be inside the unit circle
\begin{equation}
b^2 k^2  \langle \boldsymbol T_s \rangle ^2 CV^2_T+(b k \langle \boldsymbol T_s \rangle +1)^2. 
\end{equation}
It follows that in this limit, noisy transmission times may result in instability of the system. If this eigenvalue is inside the unit circle then the variance of $\bold x$ is \begin{equation} \small\begin{aligned}
\text{var}(\bold x )=\overbrace{\sigma^2 \frac{-b (CV^2_T+1)  \langle \boldsymbol T_s \rangle (3 b k \langle \boldsymbol T_s \rangle  +4)+ b^2  \langle \boldsymbol T_s^3 \rangle /\langle \boldsymbol T_s \rangle }{4 k (b k\langle \boldsymbol T_s \rangle +2)}}^\text{Fluctuations contributed from noisy channel} +\overbrace{c^2 \frac{(b k  \langle \boldsymbol T_s \rangle (3 b (CV^2_T +1) k  \langle \boldsymbol T_s \rangle +8CV^2_T)-8)- bk \langle \boldsymbol T_s^3 \rangle /\langle \boldsymbol T_s \rangle^2 }{8 b k (b k  \langle \boldsymbol T_s \rangle+2)} ,}^\text{Fluctuations contributed from disturbance}
\end{aligned}
\end{equation}
where $\langle \boldsymbol T_s^3 \rangle$ denotes the third-order moment of the time interval in between transmissions. 
While for nonzero $a$, variance of $\bold x$ depends on the entire distribution of $\boldsymbol T_s$; for $a=0$, it only depends on the first three moments of $\boldsymbol T_s$.

In order to further simplify these results, we assume that the time intervals are log-normally distributed ($\langle \boldsymbol T_s^3 \rangle= \langle \boldsymbol T_s\rangle^3 (1+CV^2_T)$) and noise in between time intervals is small, i.e. $CV^i_T \approx 0, \ i>2$. In this limit, variance of $\bold x$ simplifies to 
\begin{equation}  \begin{aligned}
\text{var}(\bold x )= & \overbrace{\sigma^2 \left( -\frac{b CV^2_T \langle \boldsymbol T_s \rangle }{k (b k \langle \boldsymbol T_s \rangle +2)}-\frac{b \langle \boldsymbol T_s \rangle}{2 k}  \right)}^\text{Fluctuations contributed from noisy channel} +\overbrace{ c^2 \left( \frac{CV^2_T  \langle \boldsymbol T_s \rangle  }{b k T+2}+ \frac{b k  \langle \boldsymbol T_s \rangle  -2}{4 b k}  \right). }^\text{Fluctuations contributed from disturbance}
\end{aligned}
\end{equation}
Given the fact that for having a finite mean one of the parameters $b$ and $k$ should be negative, then clearly in this limit, increasing mean transmission times or noise in transmission times increases variance of $\bold x$. In the next section, we illustrate our results via a two dimensional system.

\subsection{A two states NCS example}
Consider a NCS with two states $\bold x = [\bold x_1 \ \bold x_2]^\top$ that is governed by the following dynamics
\begin{equation}
d \bold x = \left(\hat{a}+A \bold x(t)+ B \bold u \right) + C d\boldsymbol{w}_2, 
\end{equation}
where 
\begin{equation}
\begin{aligned}
 \hat{a}=\left[\begin{array}{c}
a_1\\ 0
\end{array}\right] , \ \  A=  \left[\begin{array}{cc}
 -\gamma_1   & 0 \\ a_2 & -\gamma_2 
\end{array}\right] , \ \  B=\left[\begin{array}{cc}
1  & 0 \\ 0 & 1
\end{array}\right] , \ \   C=  \left[\begin{array}{cc}
\sqrt{a_1}  & 0 \\ 0 & 0
\end{array}\right]  .
\end{aligned}
\end{equation}
This example is motivated from biochemical reactors. Assume that $\bold x_1$ and $\bold x_2$ are levels of species $1$ and $2$, respectively. Then, the system under consideration can be interpreted as having production of  molecules of species $1$ at a constant rate $a_1$. The molecules species $2$ are produced from species $1$, i.e., its production rate is $a_2\bold x_1$. The production of species $\bold x_1 $ can be assumed to be noisy as the dynamics corresponds to the Langevin approximation of this reaction \cite{gil00}. Finally, molecules of $\bold x_1$ and $\bold x_2$ degrade with rates $\gamma_1$ and $\gamma_2$, respectively.

This biochemical reactor is controlled through a shared network. Any time that transmission happens, the control law changes as
\begin{equation}
\begin{aligned}
&\langle \bold{u} (\boldsymbol{t}_s^+)\rangle=  K \bold{x}(\boldsymbol{t}_s^-),\ K=  \left[\begin{array}{cc}
-k_1 &-k_2\\ 0&-k_3 
\end{array}\right] .
\end{aligned}
\end{equation}
There could be several ways to control this reactor. For example, a UV radiation that increases death rate of molecules can be used \cite{kka04, moh15c}. This control law is implemented by manipulating $k_1$ and $k_3$. Another possibility is that the resources needed to produce species $1$ are controlled based on levels of species $2$ through the parameter $k_3$. Such negative feedback loops are common motifs in biological systems \cite{bps16,sin11, svn14}.

By introducing $\bold y= [\bold x \ \bold u]^\top\in \mathbb{R}^{n+m\times 1}$, this system can be written in the form of \eqref{dynamicsSDE2}. Hence the methods explained in this paper can be applied, which results in 
	\begin{align}
\overline{\langle \bold x_1 \rangle} =\frac{a_1  (\gamma_2+k_3 )}{a_2  k_2 +\gamma_1  (\gamma_2+k_3 )+\gamma_2 k_1 +k_1 k_3 } ,  \ \  \overline{\langle \bold x_2 \rangle} =\frac{\gamma_2 +k_3 }{a_2 } \overline{\langle \bold x_1 \rangle} .\label{mean0}
\end{align}
Note that we can select $k_1$, $k_2$, and $k_3$ to have any desired mean level. Moreover, we have one degree of freedom (two means and three parameters); hence, we can use this degree of freedom to minimize one of the terms in the vector of second order moments. 
We do not show the second-order moments of $\bold x $ due to space constraints.

\section{Conclusion}

We studied statistical moments of a network control system in which transmission times in between the controller and the plant are randomly distributed. We derived exact solution of the mean and the second-order moments as well as the stability conditions. We showed that these results can be used to design controllers that maintain mean of the states at desired levels. 
We demonstrate our method on different examples. Surprisingly, as long as mean of the states of the plant is finite, mean of the state of the system is independent of transmission times statistical moments. In addition, we observed that under specific parameter regimes, rare transmissions not only save resources of the system, but may also reduces noise. Further, we showed that for a fixed mean of transmission time intervals, noisy transmission times can be used to reduce the fluctuations in $\bold x$.

Future work will extend the method explained here to network control systems in which more than just two parts are connected via a network, i.e. a system with more than one set of random transmissions. Another avenue of research would be the systems whose states switch between sub-systems, depending upon the values of the state. Finally, it would also be interesting to extend this work to analyze network control systems with nonlinearities in their dynamics. While for nonlinear systems exact solutions may not be available, moment closure methods can be used to obtain approximate analytical solutions \cite{sih10,sih10a, svs15,sih05}.

\appendix

\subsection{Proof of equation \eqref{mean of x general} }
Using \eqref{dynamicsSDE2}, the states of SHS right before $s^{th}$ event ${\bold{y}}(\boldsymbol t_{s}^-)$ is related to ${\bold{y}}(\boldsymbol t_{s-1}^+) $ as  
\begin{equation}\begin{aligned}
{\bold{y}}(\boldsymbol t_{s}^-)  = &{\rm e}^{ A_y T_s} \int_0^{T_s}  {\rm e}^{ -A_yr} \hat{a}_y dr
+  {\rm e}^{ A_y T_s } {\bold{y}}(\boldsymbol t_{s-1}^+)   
+ \int_0^{T_s}  (C_y+D_y\bold{y}(t)\mathbbm{1}_{n+m})d \boldsymbol w_{n+m} .  \label{right before}
\end{aligned}
\end{equation}
Thus the mean of the states after $s^{th}$ event is  
\begin{equation}\begin{aligned}
\langle {\bold{y}}(\boldsymbol t_{s}^+)   \rangle = &J_y \left \langle {\rm e}^{ A_y \boldsymbol T_s} \int_0^{\boldsymbol T_s}  {\rm e}^{ -A_y r} \hat{a} dr \right \rangle  + J_y \left \langle {\rm e}^{ A _y\boldsymbol T_s } \right \rangle  \left \langle  {\bold{y}}(\boldsymbol t_{s-1}^+)   \right \rangle   . 
\end{aligned}\label{xi0}
\end{equation} 
Here we used the fact that the Ito integral 
\begin{equation}\begin{aligned} \left\langle 
 \int_0^{T_s}  (C_y+D_y\bold{y}(t)\mathbbm{1}_{n+m})d \boldsymbol w_{n+m} \right \rangle  =0
\end{aligned}
\end{equation}
\cite{oks03}. Hence from \eqref{xi0}, the mean of the states right after an event in steady-state ($s\rightarrow \infty$) exists if and only and if eigenvalues of $J_y \left\langle  {\rm e}^{ A_y \boldsymbol T_s} \right \rangle$ are inside the unite circle. In this limit the steady-state mean of the states right after an event $\overline{\langle \bold y \vert_{\tau=0}\rangle }$ can be written as
\begin{equation}
\begin{aligned}
& \overline{\langle \bold y \vert_{\boldsymbol\tau=0}\rangle } =  \left(I_{n+m} -J_y \left\langle {\rm e}^{ A_y \boldsymbol T_s} \right \rangle   \right)^{-1}J_y \left \langle  {\rm e}^{ A_y \boldsymbol T_s} \int_0^{\boldsymbol T_s}  {\rm e}^{ -A_y r} \hat{a}_y  dr  \right \rangle  .
\end{aligned}
\label{mean of xtau0}	
\end{equation}

By using equation \eqref{mean of xtau0}, the mean of the states in between events for any given $\tau$ is  
\begin{equation}
\begin{aligned}
\overline{\langle y \vert_{\boldsymbol\tau=\tau }\rangle } = & e^{A_y\tau } \left(I_{n+m} -J_y \left\langle {\rm e}^{ A_y \boldsymbol T_s} \right \rangle   \right)^{-1} J\left \langle  {\rm e}^{ A_y\boldsymbol T_s} \int_0^{\boldsymbol T_s}  {\rm e}^{ -A_yr} \hat{a}_y dr  \right \rangle   +  {\rm e}^{ A_y \tau  } \int_0^\tau   {\rm e}^{ -A_yr} \hat{a}_y dr  .
\label{first x}
\end{aligned}
\end{equation}
The mean of the states is obtained by unconditioning \eqref{first x} with respect to $\boldsymbol \tau$ using \eqref{prob. tau} as shown in \eqref{mean of x general}.

\subsection{Dynamics of second-order moments of NCS}
We start by calculating dynamics of $\bold x \bold x^\top$ in between random transmissions.
Using \eqref{dynamicsSDE}, the time evolution of $\bold x \bold x^\top$ is obtained as 
\begin{equation}
\begin{aligned}
\frac{d \bold{x}\bold{x}^\top}{dt}  
=    A  \bold{x}\bold{x}^\top +    \bold{x}\bold{x}^\top  A^\top  +B   \bold{u}\bold{x}^\top + \bold{x}\bold{u}^\top  B^\top   +\hat{a}\bold{x}^\top   +  \bold{x}   \hat{a}^\top +CC^\top  + C \mathbbm{1}_{n}^\top  \bold{x}^\top  D^\top + D   \bold{x}  \mathbbm{1}_{n} C +nD  \bold{x}\bold{x}^\top D^\top .
\label{second order dunamics}
\end{aligned}
\end{equation}
By vectorization, we can rewrite \eqref{second order dunamics} as
\begin{equation}\begin{aligned}
\frac{d {\rm vec}  \left( \bold x  \bold x^\top \right) }{dt } = &  \left(I_n \otimes A+  A \otimes I_n +n D\otimes D\right)   
{\rm vec}  \left( \bold x  \bold x^\top\right)  
+\left(I_n \otimes B+  B \otimes I_n \right)   
{\rm vec}  \left( \bold x  \bold u^\top\right)  + {\rm vec}  \left( CC^\top\right)  \\
& +( I _n\otimes \hat{a} + \hat{a}  \otimes I_n+D \otimes C  \mathbbm{1}_{n}+ C^\top  \mathbbm{1}_{n} \otimes D)\bold x,
\end{aligned}
\label{kronecker x2}
\end{equation} 
where we used the fact that for three matrices $M$, $M'$, and $M''$
\begin{equation}
{\rm vec}(M M' M'') = (M''^\top \otimes M){\rm vec}(M') 
\end{equation}
\cite{mao13}. Since these dynamics depend on $\bold x  \bold u^\top$ we also add dynamics of  $\bold x  \bold u^\top$
\begin{equation}
\begin{aligned}
\frac{d \bold{x}\bold{u}^\top}{dt}  
= &   A  \bold{x}\bold{u}^\top+ B  \bold{u}  \bold{u}^\top  +  \hat{a} \bold{u}^\top  .
\label{second order dunamics000}
\end{aligned}
\end{equation}
Again by using vectorization we have 
\begin{equation}
\begin{aligned}
\frac{d {\rm vec}  \left( \bold x  \bold u^\top \right) }{dt } =& ( I_m \otimes A ) {\rm vec}  \left( \bold x  \bold u^\top\right) + ( I_m \otimes B) {\rm vec}  \left( \bold u  \bold u^\top\right)  + ( I_m \otimes \hat{a} )  \bold u
\label{second order dunamics00}
\end{aligned}
\end{equation}
and finally we add dynamics of ${\rm vec}  \left( \bold u  \bold u^\top \right)$ by using \eqref{control law}
\begin{equation}
\begin{aligned}
&\frac{d {\rm vec}  \left( \bold u  \bold u^\top \right) }{dt } =0.
\label{second order dunamics0}
\end{aligned}
\end{equation}
The combination of \eqref{kronecker x2}, \eqref{second order dunamics00}, and \eqref{second order dunamics0} results in \eqref{mu dynamic0}.

\subsection{Second-order moments of NCS after transmissions}
Based on \eqref{reset} 
\begin{equation}\begin{aligned}
\langle & \bold u \bold   u^\top (\boldsymbol{t}_s^+)\rangle =   K \bold x \bold x^\top (\boldsymbol{t}_s^-)  K^\top+ \Sigma \Rightarrow 
{\rm vec}  \langle \bold u \bold u^\top (\boldsymbol{t}_s^+)\rangle  = K \otimes K {\rm vec}  ( \bold x \bold x^\top (\boldsymbol{t}_s^-))+ {\rm vec}  (\Sigma).
\end{aligned}
\end{equation}
Further $\langle \bold x \bold u (\boldsymbol{t}_s^+)\rangle $ can be written as
\begin{equation}\begin{aligned}
\langle \bold x \bold u^\top (\boldsymbol{t}_s^+)\rangle & =   \bold x \bold x^\top (\boldsymbol{t}_s^+)  K^\top\Rightarrow 
 {\rm vec}  \langle \bold x \bold u^\top (\boldsymbol{t}_s^+)\rangle  = K \otimes I_n {\rm vec}  ( \bold x \bold u^\top (\boldsymbol{t}_s^-)).
\end{aligned}
\end{equation}
Overall we have
\begin{equation}
\begin{aligned}
\langle \boldsymbol \mu(\boldsymbol{t}_s^+)  \rangle =   J_\mu \boldsymbol \mu(\boldsymbol{t}_s^-) + R_\mu. \label{reset 20}
\end{aligned}
\end{equation}
where 
\begin{equation}
J_\mu \equiv    \left[\begin{array}{c;{2pt/2pt}c;{2pt/2pt}c;{2pt/2pt}c;{2pt/2pt}c}
I_n & 0 & 0 & 0& 0\\ \hdashline[2pt/2pt]0 &  K & 0 & 0&0 \\ \hdashline[2pt/2pt] 0& 0& I_{n+m}&0& 0  \\ \hdashline[2pt/2pt] 0& 0&  K \otimes I_n &0 & 0\\ \hdashline[2pt/2pt] 0&0&  K \otimes K  &0& 0	\end{array}\right],
R_\mu \equiv  \left[\begin{array}{c}
\hat{a}\\ \hdashline[2pt/2pt]0\\ \hdashline[2pt/2pt] 0\\ \hdashline[2pt/2pt]0\\ \hdashline[2pt/2pt]{\rm vex}\left(\Sigma\right) 
\end{array}\right]. \label{Jmu}
\end{equation}
Note that in deriving ${J_\mu}$ we used the fact that values of $\bold x$ before and after transmission are equal hence
\begin{equation}\begin{aligned}
\langle  \bold x \bold   x^\top (\boldsymbol{t}_s^+)\rangle - \langle \bold x (\boldsymbol{t}_s^+) \rangle \langle  \bold   x^\top (\boldsymbol{t}_s^+)\rangle =0  \Rightarrow 
{\rm vec}  \langle \bold x \bold x^\top (\boldsymbol{t}_s^+)\rangle  =  {\rm vec}  ( \bold x (\boldsymbol{t}_s^-) \bold x^\top (\boldsymbol{t}_s^-)).
\end{aligned}
\end{equation}

\bibliographystyle{plos2009}
\bibliography{RefMaster}

\end{document}